\documentclass[conference]{IEEEtran}
\IEEEoverridecommandlockouts
\usepackage{blindtext, graphicx, geometry}
\usepackage{multirow, array, mathtools}

\usepackage{mathtools}
\usepackage{cite}
\usepackage{float}
\usepackage{multicol}
\usepackage{multirow}
\usepackage{hyperref}
\usepackage{graphicx}
\usepackage{amsfonts}
\usepackage{amsmath}
\usepackage{color}
\usepackage{comment}
\usepackage{epstopdf}
\usepackage{subcaption}
\usepackage{mathtools, nccmath}
\usepackage{multirow}
\usepackage{bbm}
\usepackage{marginnote}
\usepackage{tikz}
\usepackage{bm}
\usetikzlibrary{arrows,decorations.pathmorphing,backgrounds,fit,positioning,shapes.symbols,chains, shapes,snakes}
\usepackage{eurosym}
\usepackage{comment}
\usepackage{makecell}

\usepackage[font={small}]{caption}

\ifCLASSINFOpdf
\else
\fi

\begin{document}

\title{Toward Coordinated Transmission and Distribution Operations \vspace{-0.5cm}}

\author{Mikhail Bragin, \textit{IEEE, Member}, Yury Dvorkin, \textit{IEEE, Member}. \vspace{-0.5cm} }

\maketitle

\begin{abstract}
Proliferation of smart grid technologies has enhanced observability and controllability of distribution systems. If coordinated with the transmission system, resources of both systems can be used more efficiently. This paper proposes a model to operate transmission and distribution systems in a coordinated manner. The proposed model is solved using a Surrogate Lagrangian Relaxation (SLR) approach. The computational performance of this approach is compared  against existing methods (e.g. subgradient method). Finally, the usefulness of the proposed model and solution approach is demonstrated via numerical experiments on the illustrative example and IEEE benchmarks. 

\end{abstract}



%
\IEEEpeerreviewmaketitle
\vspace{-3mm}
\section{Introduction}
\vspace{-1.5mm}
Active deployment of smart grid technologies in distribution systems has affected the way how these systems interact with the transmission systems. It is anticipated that distribution systems of the future will be equipped to actively engage in transmission system operations, \cite{TD1}. This  will require a coordination mechanism to co-optimize generation resources available in both systems to achieve least-cost operations, while respecting objective functions and satisfying technical constraints of each system. 

Coordination between the transmission and distribution systems has previously been investigated for economic dispatch and optimal power flow frameworks.  References \cite{TD2, TD3} propose a decomposition approach for the coordinated economic dispatch of the transmission and distribution systems that can capture heterogeneous technical characteristics of these systems. In \cite{TD4}, the decomposition algorithm from \cite{TD2, TD3}  is improved to handle AC power flow constraints for both the transmission and distribution systems. The interactions between the transmission and distribution system in the electricity market context is studied in \cite{Caramanis_2015}. The common caveat of  \cite{TD2, TD3, TD4, Caramanis_2015} is that they do not endogenously model binary unit commitment (UC) decisions on conventional generators. To our knowledge, there is no approach to include binary UC decisions, while coordinating transmission and distribution operations. 

Considering binary UC decisions invokes a number of challenges. First, it renders  a mixed-integer linear program (MILP) that cannot always be solved efficiently with a standard branch-and-cut method. Second, traditional decomposition techniques, e.g. Lagrangian Relaxation (LR), are notorious for their unstable and often slow convergence due to the zigzagging effect of Lagrange multipliers. This paper deals with both challenges by using the Surrogate Lagrangian Relaxation (SLR) \cite{TD5}. The SLR enforces a ``surrogate optimality'' condition, which guarantees that ``surrogate'' subgradient directions form acute angles with directions toward the optimal multipliers. The ``surrogate optimality'' condition makes it unnecessary to solve all decomposed subproblems to optimality, thus speeding up the computations. Reference \cite{TD5} derives a stepsizing formula that guarantees the convergence and quantifiable solution accuracy without requiring any knowledge of the optimal dual Lagrangian function. Previously, the SLR was applied to large-scale transmission UC models \cite{TD6, TD9}, even with  AC power flows \cite{TD10}.  

This paper proposes  a model to coordinate the transmission and distribution systems, which  accounts for binary UC decisions and power flow physics. The model is solved using the SLR. Our case study describes the cost and computational performance of the proposed coordination and solution technique.
\vspace{-7pt}
\section{Model}
\vspace{-3pt}
This paper considers a power system layout typical to the US power sector, where multiple distribution systems are connected to the single transmission system. The transmission system is operated by the transmission system operator (TSO) using a wholesale electricity market. Each distribution system is  operated by the distribution system operator (DSO) that dispatches its own generation and can also  participate in the wholesale electricity market. 

\vspace{-7pt}

\subsection{Preliminaries} \label{sec:prelim}
Let $\mathcal{B}^{[\cdot]}$, $\mathcal{I}^{[\cdot]}$ and $\mathcal{L}^{[\cdot]}$ be the sets of buses, generators and lines indexed by $b$, $i$, and $l$, where superscript ${[\cdot]}$ denotes the transmission ($\text{T}$) and distribution ($\text{D}$) system. Let $\mathcal{J}$ be the set of distribution systems indexed by $j$. 
The transmission system and each distribution system are then given by graphs $\mathcal{G}^{\text{T}} = (\mathcal{B}^{\text{T}}, \mathcal{L}^{\text{T}})$ and $\mathcal{G}^{\text{D}}_j = (\mathcal{B}^{\text{D}}_j, \mathcal{L}^{\text{D}}_j)$.  Graph $\mathcal{G}^{\text{T}}$  is chosen to be loopy (meshed) and $\mathcal{G}^{\text{D}}_j$ is chosen to be tree (radial) to represent  common topologies of the transmission and distribution systems. Graph $\mathcal{G}^{\text{T}}$ and each graph $\mathcal{G}^{\text{D}}_j$  have strictly one connection point at the root bus  of $\mathcal{G}^{\text{D}}_j$. The root bus of each distribution system is denoted as $b_{0,j}$. To denote the connection between the transmission and distribution systems, we use index $j(b)$, which is interpreted as distribution system $j$ is connected to transmission bus $b$. The set of transmission buses that have distribution systems is denoted as $\hat{B}^{\text{T}}$. Active and reactive power variables are distinguished by superscripts $\text{p}$ and $\text{q}$.

\subsection{DSO Model}\label{sec:dso}
The following model is formulated for each distribution system individually and therefore index $j$ is omitted for the sake of clarity. The DSO aims to  maximize the social welfare in the distribution system by supplying its demand using available distribution and wholesale market resources:

\begin{flalign}
\max \!\! \big\{ o^{\text{D}} \big\} \!\! =\!\! \max \!\bigg\{\!\sum_{b \in \mathcal{B}}^{} \!L_b^{\text{p}} T\!\!-\!\!\sum_{i \in \mathcal{I}^{\text{D}}} C_i^{\text{g}} g_i^{\text{p}}\! \!+\! \lambda_{b_0} (p^{\text{$\uparrow$}}_{b_0} \!-\! p^{\text{$\downarrow$}}_{b_0})\!\! \bigg\}. \label{dso_obj}
\end{flalign}
The first term in \eqref{dso_obj} represents the payment collected by the DSO from consumers based on their active power consumption $L_b^{\text{p}}$  and flat-rate tariff $T$. The second term accounts for the production cost of conventional generators located in the distribution system and is computed based on their incremental generation cost $C_i^{\text{g}}$   and active power output $g_i^{\text{p}}$. The third term accounts for the cost of transactions performed by the DSO in the wholesale electricity market. Variables $p^{\text{$\downarrow$}}_{b_0}$  and $p^{\text{$\uparrow$}}_{b_0}$  represent the capacity bid/offered by the DSO in the wholesale market, while $\lambda_{b_0}$   denotes the locational marginal price (LMP) at the transmission bus, which is connected to the root bus of the distribution system. Thus, $p^{\text{$\uparrow$}}_{b_0} > 0$ indicates that the DSO offers to sell electricity in the wholesale market, while $p^{\text{$\downarrow$}}_{b_0} > 0$ signals that the DSO bids to purchase electricity 
Note \eqref{dso_obj} neglects the fixed  cost of conventional generators as it is normally negligible for distribution generators.  

The output of distribution generators is constrained as:
\begin{flalign}
&  \underline{G}^{\text{p}}_i \leq g^{\text{p}}_i \leq \overline{G}^{\text{p}}_i, \quad \forall i \in \mathcal{I}^{\text{D}}, \label{dso_eq1} \\
& \underline{G}^{\text{q}}_i \leq  g^{\text{q}}_i \leq \overline{G}^{\text{q}}_i, \quad \forall i \in \mathcal{I}^{\text{D}},\label{dso_eq3}
\end{flalign}
where the minimum an maximum active power limits are $\underline{G}^{\text{p}}_i$ and $\overline{G}^{\text{p}}_i$, while the minimum and maximum reactive power limits are $\underline{G}^{\text{q}}_i$ and $\overline{G}^{\text{q}}_i$. sSince this paper considers a single-period optimization, the economic dispatch constraints do not include inter-temporal limits (e.g. ramp limits).  

Since the distribution system is assumed to have a radial topology, AC power flows can be modeled using an exact second-order conic (SOC) relaxation; interested readers are referred to \cite{low_relaxation} for details of this relaxation given below:
\begin{flalign}
& \big[(f^{\text{p}}_l)^2 + (f^{\text{q}}_l)^2\big]\frac{1}{a_l} \leq v_{s(l)}, \quad \forall l \in \mathcal{L}^{\text{D}},  \label{dso_eq8} \\
& v_{r(l)} - 
v_{s(l)} = 2 (R_l f^{\text{p}}_l + X_l f^{\text{q}}_l) - a_l (R_l^2 + X_l^2),  \quad \forall l \in \mathcal{L}^{\text{D}}, \label{dso_eq7} \\
& (f^{\text{p}}_l)^2 + (f^{\text{q}}_l)^2 \leq \overline{S}_l^2, \quad \forall l \in \mathcal{L}^{\text{D}},  \label{dso_eq5} \\
& (f^{\text{p}}_l - a_l R_l)^2 +  (f^{\text{q}}_l - a_l X_l)^2  \leq \overline{S}_l^2, \quad \forall l \in \mathcal{L}^{\text{D}}  \label{dso_eq6} \\
& \underline{V}_b \leq v_b \leq \overline{V}_b, \quad \forall b \in \mathcal{B}^{\text{D}}. \label{dso_eq16} 
\end{flalign}
Eq.~\eqref{dso_eq8} represents a relaxed expression for the current squared in branch $l$, denoted by auxiliary variable $a_l$,  variables $f^{\text{p}}_l$  and $f^{\text{q}}_l$ denote active and reactive power flows across line $l$, and $v_{s(l)}$ is the voltage magnitude at the sending end of line $l$. The sending and receiving buses of branch $l$ are denoted as $s(l)$  and $r(l)$, respectively. Eq.~\eqref{dso_eq7} relates the sending and receiving bus voltages squared  $v_{s(l)}$ and $v_{r(l)}$ via the voltage drop across branch $l$, where parameters $R_l$ and $X_l$ are the reactanace and impedance of branch $l$. Since the power flow at the sending and receiving buses of each branch $l$ differs due to losses incurred by transmission, the apparent power flow limit $\overline{S}_{l}$  is enforced for the sending and receiving buses separately in \eqref{dso_eq5} and \eqref{dso_eq6}.  The bus voltages are constrained  in \eqref{dso_eq16}, where  $v_b$  denotes voltages squared limited by $\underline{V}_b$ and $\overline{V}_b$, see \cite{low_relaxation}.

With the exception of the root bus, which is discussed below, the nodal power balance is enforced as:
\begin{flalign}
& f_{l|s(l)=b}^{\text{p}} - \sum_{l|r(l)=b  } (f_l^{\text{p}} - a_l R_l) - \sum_{i \in \mathcal{I}_b^{\text{U}}}g_i^{\text{p}} + L_b^{\text{p}} + v_b  G_{l|s(l)=b} \nonumber \\ &  \hspace{4.2cm} =0,  \quad  \forall b \in \mathcal{B}^{\text{D}} \backslash \big\{ b_0 \big\},\label{dso_eq9} \\
& f_{l|s(l)=b}^{\text{q}} -\!\! \sum_{l|r(l)=b  } \!\!(f_l^{\text{q}} - a_l X_l) - \sum_{i \in \mathcal{I}_b^{\text{U}}}g_i^{\text{q}} + L_b^{\text{q}} - v_b  B_{l|s(l)=b} \nonumber \\ &  \hspace{4.2cm}  =0,  \quad\forall b \in \mathcal{B}^{\text{D}} \backslash \big\{ b_0 \big\} \label{dso_eq10}, 
\end{flalign}
where $L_b^{\text{p}}$ and $L_b^{\text{q}}$ denote the active and reactive power consumption at bus $b$ and $G_l$ is the conductance of branch $l$. In case of the root bus, \eqref{dso_eq9} and \eqref{dso_eq10} transform into:
\begin{flalign}
&  \!\!- \!\! \sum_{l|r(l)=b_0  } \!\!\!\!(f_l^{\text{p}} - a_l R_l) \!- \!p^{\uparrow}_{b_0} + p^{\downarrow}_{b_0} + v_{b_0}  G_{l|o(l)=b_0} =0, \label{dso_eq11} \\
&  \!\!- \!\!\sum_{l|r(l)=b_0  } \!\!(f_l^{\text{q}} - a_l X_l) - v_{b_0}  G_{l|o(l)=b_0}  =0 \label{dso_eq12}. 
\end{flalign}
Eq.~\eqref{dso_eq11} includes the power exchange with the transmission system based on the capacity bid ($p^{\text{$\uparrow$}}_{b_0}$) and offered ($p^{\text{$\downarrow$}}_{b_0}$) by the DSO in the electricity market. Since the DSO is assumed to meet its own reactive power needs, the reactive power balance for the root bus in \eqref{dso_eq12} has no reactive power exchange with the transmission system. Since the physical interface between the transmission and distribution systems is limited, $p^{\text{$\downarrow$}}_{b_0}$ and $p^{\text{$\uparrow$}}_{b_0}$ are limited as:
\begin{flalign}
&  0 \leq p^{\uparrow}_{b_0} \leq \overline{P}_{j(b)}, \label{dso_eq13}\\
&  0 \leq p^{\downarrow}_{b_0} \leq \overline{P}_{j(b)} \label{dso_eq14},
\end{flalign}
where $\overline{P}_{j(b)}$ and $\overline{P}_{j(b)}$ is the active power limit between distribution system $j$ and  transmission bus $b$.

\subsection{TSO Model} \label{sec:tso}

As in \eqref{dso_obj}, the TSO aims to maximize the social welfare in the transmission system, which can be formalized as:

\begin{flalign}
& \max \big\{ o^{\text{T}} \big\} =  \bigg\{   \sum_{b \in \mathcal{B}^{\text{T}}}  C^{\text{b}}_b L^{\text{p}}_b-  \sum_{i \in \mathcal{I}^{\text{T}}} C^{\text{o}}_i  g^{\text{p}}_i   \label{tso_obj} \\  & \hspace{3.2cm}  +\sum_{ b \in \hat{\mathcal{B}^{\text{T}}}  }\bigg( C_{j(b)}^{\text{$\downarrow$}} p^{\text{$\downarrow$}}_{j(b)}  - C^{\text{$\uparrow$}}_{j(b)} p^{\text{$\uparrow$}}_{j(b)} \bigg) \bigg\} \nonumber . 
\end{flalign}
The first term in \eqref{tso_obj} represents the payment collected from consumers connected directly to the transmission system based on their active power consumption $L_b^{\text{p}}$  and price bids $C^{\text{b}}_{b}$. The second term represents the cost of offers by conventional generation resources computed based on their offered price $C^{\text{o}}_i$ and power production  $g^{\text{p}}_i $. The third term is the cost of active power exchange between the TSO and DSO, where $C^{\text{$\downarrow$}}_{j(b)}$ and $C^{\text{$\uparrow$}}_{j(b)}$ are the price bids and offers of the DSO $j$ located at transmission bus $b$. 

The dispatch of conventional generators is  constrained as:
\begin{flalign}
& \underline{G}^{\text{p}}_i \leq  g^{\text{p}}_i \leq \overline{G}^{\text{p}}_i x_i,  \quad  \forall i \in \mathcal{I}^{\text{T}}, \label{tso_eq2} 
\end{flalign}
where $x_i \in \big\{ 0, 1\big\}$ is a binary (on/off) decision on conventional generators.  Since this paper considers a single-period case, inter-temporal  ramp limits and minimum up an down times of conventional generators  are omitted.  

The network constraints are modeled using the DC power flow approximation to account for a meshed topology as customarily used in market clearing procedures:
\begin{flalign}
& f^{\text{p}}_l = \frac{1}{X_l} ( \theta_{o(l)} - \theta_{r(l)}) ,  \quad  \forall l \in \mathcal{L}^{\text{T}}, \label{tso_eq6} \\
& -\overline{F}_l \leq f^{\text{p}}_l \leq \overline{F}_l  , \quad \forall l \in \mathcal{L}^{\text{T}}, \label{tso_eq7}
\end{flalign}
where \eqref{tso_eq6} computes the active power flow in line $l$ and the active power flow limit $\overline{F}_l$ on each line $l$ is enforced in  \eqref{tso_eq7}. The nodal active power balance is then modeled for transmission buses without and with interconnected distribution systems in \eqref{tso_eq1b} and \eqref{tso_eq1}:
\begin{flalign}
& \sum_{i \in I_b} g^{\text{p}}_i   + \sum_{l|r(l)=b} f_l^{\text{p}} - \sum_{l|o(l)=b} f^{\text{p}}_l -
 L^{\text{p}}_b =0, \nonumber \\ 
&  \hspace{4.9cm} \quad  \forall b \in \mathcal{B}^{\text{T}} \backslash \big\{ \hat{\mathcal{B}^{\text{T}}}   \big\},  \label{tso_eq1b}  \\
& \sum_{i \in \mathcal{I}_{b}} g^{\text{p}}_i   + \sum_{l|r(l)=b} f_l^{\text{p}} - \sum_{l|o(l)=b} f^{\text{p}}_l + p^{\text{$\uparrow$}}_{j(b)}- p^{\downarrow}_{j(b)} - 
 L^{\text{p}}_{b} \nonumber \\ &  \hspace{4.9cm} = 0, \forall b \in \hat{\mathcal{B}^{\text{T}}}   : (\lambda_{b}) \label{tso_eq1},
\end{flalign}
 where $\lambda_{b}$ is a Lagrangian multiplier of the power balance constraint, i.e. the wholesale LMP, at the transmission bus with an interconnected distribution system.   Variables $p^{\text{$\uparrow$}}_{j(b)}$ and $p^{\text{$\downarrow$}}_{j(b)}$ in \eqref{tso_eq1} denote the power exchnage with distribution system as seen from the transmission side. Therefore, as in  \eqref{dso_eq13}-\eqref{dso_eq14}, these flows are constrained:
\begin{flalign}
&  0 \leq p^{\downarrow}_{j(b)} \leq \overline{P}_{j(b)}, \quad \forall b \in \hat{\mathcal{B}^{\text{T}}},    \label{tso_eq4}\\
&  0 \leq p^{\uparrow}_{j(b)} \leq \overline{P}_{j(b)}, \quad  \forall b \in \hat{\mathcal{B}^{\text{T}}}.      \label{tso_eq5}
\end{flalign}

\subsection{Coordinated TSO-DSO Model} \label{sec:coordinated}
Operating decisions of the TSO and multiple DSOs can be coordinated by solving the following problem:
\begin{flalign}
& \text{Eq.~\eqref{dso_obj}-\eqref{dso_eq14}}, \quad \forall j \in \mathcal{J} , \label{coord_eq1}\\
& \text{Eq.~\eqref{tso_obj}-\eqref{tso_eq5}},  \label{coord_eq2} \\
& p^{\text{$\downarrow$}}_{b_{0,j(b)}} = p^{\text{$\downarrow$}}_{j(b)}, \quad \forall   b \in \hat{\mathcal{B}^{\text{T}}} : (\psi_{b_{0,j}}^{\downarrow}),   \label{tso_eq5aa} \\
& p^{\text{$\uparrow$}}_{b_{0,j(b)}} = p^{\text{$\uparrow$}}_{j(b)}, \quad   \forall b \in \hat{\mathcal{B}^{\text{T}}} : (\psi_{b_{0,j}}^{\uparrow}).       \label{tso_eq5bb}
\end{flalign}
Eq.~\eqref{coord_eq1} and \eqref{coord_eq2} list all DSO and TSO problems, while \eqref{tso_eq5aa} and \eqref{tso_eq5bb} enforce the power exchanges between the DSO and TSO problems. Note that $\psi_{b_{0,j}}^{\downarrow}$ and $\psi_{b_{0,j}}^{\uparrow}$ denote Lagrange multipliers of respective constraints.  The problem in \eqref{coord_eq1}-\eqref{tso_eq5bb} cannot be solved efficiently for large-scale instances using off-the-shelf solution strategies. Furthermore, it is important to preserve the distributed nature of the coordination process between the TSO and DSOs. This motivates an iterative SLR-based solution technique described in Section~\ref{sec:solution_technique}.

\section{Solution Technique}\label{sec:solution_technique}

\begin{figure}[!b]
\centering
\begin{tikzpicture}[auto, node distance=2cm,auto, font=\normalsize]
\tikzstyle{startstop} = [rectangle, rounded corners, minimum width=3cm, minimum height=1cm,text centered, draw=black, fill=white!30]
\tikzstyle{process} = [rectangle, minimum width=3cm, minimum height=1cm, text centered, draw=black, fill=white!30]
\tikzstyle{decision} = [diamond, minimum width=2.5cm, minimum height=1cm, text centered, draw=black, fill=white!30]

\node (start) [startstop] {Initialize $\lambda^0_{b_{0,j}}, \psi^{\uparrow,0}_{b_{0,j}}, \psi^{\downarrow,0}_{b_{0,j}}, s^0, c^0, p^{0,\downarrow}_{b_{0,j}}, p^{0,\uparrow}_{b_{0,j}}, p^{0,\downarrow}_{j(b)}, p^{0,\uparrow}_{j(b)}$};
\node (DSO) [process, below of=start] {Solve the DSO problem, eq.~\eqref{alg_DSO_obj}-\eqref{alg_DSO_const}};
\node (TSO) [process, below of=DSO] {Solve the TSO problem, eq.~\eqref{relaxation_eq1}-\eqref{eq_opt}};
\node (Update) [process, below of=TSO] {Update $\lambda^k_{b_{0,j}}, \psi^{\uparrow,k}_{b_{0,j}}, \psi^{\downarrow,k}_{b_{0,j}}, s^k, c^k, \alpha^k$};
\node (Termination) [decision, below of=Update, node distance=2cm] {Stop?};
\node (end) [startstop, below of=Termination] {End};

    \coordinate[right of=DSO, xshift=1.5cm] (c2);   
    \coordinate[right of=Termination, xshift=1.5cm] (c1);  
    
    \draw [->,thick, line width=0.25mm] (start) -- (DSO) node [midway] {};
    \draw [->,thick, line width=0.25mm] (DSO) -- (TSO) node [midway] {$ p^{\uparrow,k}_{b_{0,j}}, p^{\downarrow,k}_{b_{0,j}}$};
    \draw [->,thick, line width=0.25mm] (TSO) -- (Update) node [midway] {};    
    \draw [->,thick, line width=0.25mm] (TSO) -- (Update) node [midway] {$f^k_l, v_b^k, a_l^k, \theta_b^k$};
    \draw [-,thick, line width=0.25mm] (Update) -- (TSO) node [midway] {$p^{\uparrow,k}_{b_{0,j}}, p^{\downarrow,k}_{b_{0,j}}, p^{\uparrow,k}_{j(b)}, p^{\downarrow,k}_{j(b)}$};    
    \draw [->,thick, line width=0.25mm] (Update) -- (Termination) node [midway] {};  
    \draw [->,thick, line width=0.25mm] (Termination) -- (end) node [midway] {}; 
    \draw [-,thick, line width=0.25mm]  (Termination) -- (c1) node [midway] {};  
    \draw [-,thick, line width=0.25mm]  (c1) -- (c2) node [midway, right, xshift=0.25cm, yshift=-0.5cm, rotate=90] {$k=k+1$};
    \draw [->,thick, line width=0.25mm] (c2) -- (DSO) node [midway] {}; 
\end{tikzpicture}
\caption{Flowchart of the proposed SLR-based solution technique. }
\label{fig_algortihm}
\end{figure}
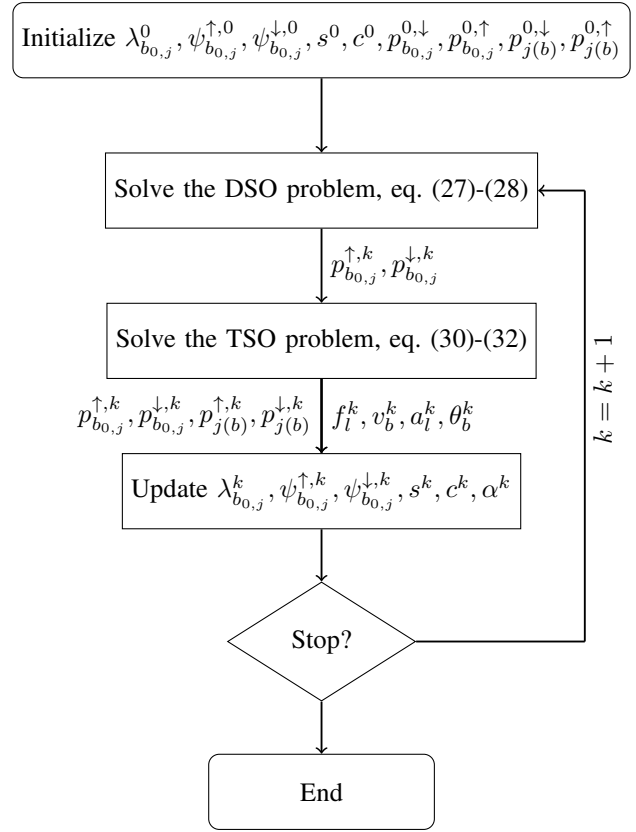

The proposed SLR-based solution technique is illustrated in Fig.~\ref{fig_algortihm} and each step is detailed below:

\subsubsection{Initialization} Set the iteration counter $k=0$. Stepsize $s^{0}$ are initialized as in \cite{TD5} and penalty coefficient $c^{0}$ is  chosen  as in\cite{bertsekas}. Also, initialize  $\lambda^{0}_{b_0,j}, \psi^{0,\downarrow}_{b_{0,j}}, \psi^{0,\uparrow}_{b_{0,j}}, p^{0,\downarrow}_{b_{0,j}}, p^{0,\uparrow}_{b_{0,j}}, p^{0,\downarrow}_{j(b)}, p^{0,\uparrow}_{j(b)}$.
\subsubsection{Solve the DSO problem} The following problem is solved for each DSO in a parallel manner ($\forall j \in \mathcal{J}$) 

\begin{flalign}
& \max o^{\text{D}}_j (g_i, p_{b_{0,j}}^{\uparrow},p_{b_{0,j}}^{\downarrow}) + \psi^{k,\downarrow}_{b_{0,j}} (p^{\downarrow}_{b_{0,j}} - p^{\downarrow,k-1}_{j(b)})  \nonumber   \\& +\frac{c^{k}}{2} |p^{\downarrow,k-1}_{b_{0,j}} - p^{\downarrow,k-1}_{j(b)}||p^{\downarrow}_{b_{0,j}} - p^{\downarrow,k-1}_{j(b)}| 
+ \psi^{k,\uparrow}_{b_{0,j}} (p^{\uparrow}_{b_{0,j}} - p^{\uparrow,k-1}_{j(b)})   \nonumber \\& +\frac{c^{k}}{2} |p^{\uparrow,k-1}_{b_{0,j}} - p^{\uparrow,k-1}_{j(b)}||p^{\uparrow}_{b_{0,j}} - p^{\uparrow,k-1}_{j(b)}|, \label{alg_DSO_obj} \\
& \text{Eq.}~\eqref{dso_obj}-\eqref{dso_eq14} \label{alg_DSO_const}. 
\end{flalign}
Since \eqref{tso_eq5aa}-\eqref{tso_eq5bb} are relaxed, the deviations from the TSO  power flows at the previous iterations, $p^{\downarrow,k-1}_{j(b)}$ and $p^{\uparrow,k-1}_{j(b)}$, are penalized in \eqref{alg_DSO_obj}.  As in \cite{TD9}, the absolute value penalties are used to avoid unnecessary linearization. 

\subsubsection{Solve the TSO problem} Following the DSO problems, optimized values of $p^{\uparrow,k}_{b_{0,j}}, p^{\downarrow,k}_{b_{0,j}}$ are used in the TSO problem: 

\begin{flalign}
& \max \big\{ 
L_{c^{k}}(\lambda^{k}_{b_{0,j}}, \psi^{\uparrow,k}_{b_{0,j}}, \psi^{\downarrow,k}_{b_{0,j}}; p^{\uparrow,k}_{b_{0,j}}, p^{\downarrow,k}_{b_{0,j}};  \\ & \hspace{5cm}  f_l, g_l, \theta_b, p_{j(b)}^{\uparrow},p_{j(b)}^{\downarrow}) 
\big\},\label{relaxation_eq1} \\
&  \text{Eq.}~\eqref{tso_eq2}-\eqref{tso_eq1b},~\eqref{tso_eq4}-\eqref{tso_eq5},  \label{relaxation_eq2} \\
& \tilde{L}_{c^{k}}(\lambda^{k}_{b_{0,j}}, \psi^{\uparrow,k}_{b_{0,j}}, \psi^{\downarrow,k}_{b_{0,j}}; p^{\uparrow,k}_{b_{0,j}}, p^{\downarrow,k}_{b_{0,j}};  f_l^{k}, g_l^{k}, \theta_b^{k}, p_{j(b)}^{\uparrow,k}, 
 p_{j(b)}^{\downarrow,k}) > \nonumber \\ &  \tilde{L}_{c^{k}}(\lambda^{k}_{b_{0,j}}, \psi^{\uparrow,k}_{b_{0,j}}, \psi^{\downarrow,k}_{b_{0,j}}; p^{\uparrow,k}_{b_{0,j}}, p^{\downarrow,k}_{b_{0,j}}; f_l^{k-1},   g_l^{k-1}, \theta_b^{k-1},  \nonumber \\ & \hspace{5cm}  p_{j(b)}^{\uparrow,k-1},p_{j(b)}^{\downarrow,k-1}), \label{eq_opt}
 \end{flalign}
where $L_{c^{k}}$ is the augmented Lagrangian function, and $\tilde{L}_{c^{k}}$ is the surrogate augmented dual value. The value of $\tilde{L}_{c^{k}}$ is defined as the value of  (29) for its current feasible solution. Eq.~\eqref{eq_opt} represents the ``surrogate optimality'' condition from  \cite{TD9}. As in Step 2, we relax and penalize  constraints (20) and (25)-(26) within the  augmented Lagrangian function $L_{c^{k}}$. The penalization is implemented as discussed in \cite{TD9}.  Due to the pagination limit, we omit  the procedure to derive the exact expressions for $L_{c^{k}}$  and $\tilde{L}_{c^{k}}$ and refer interested readers to \cite{TD9} for details.



\subsubsection{Update} Using the DSO and TSO solutions obtained at iteration $k$, the following parameters are updated:
 \begin{flalign}
& c^{k+1} =  c^k\beta, \beta > 1, \\
& \psi_{b_{0,j}}^{\downarrow,k+1} =  \psi_{b_{0,j}}^{\downarrow,k} +s^k (p^{\downarrow,k}_{b_{0,j}} - p^{\downarrow,k}_{j(b)}), \\
& \psi_{b_{0,j}}^{\uparrow,k+1} =  \psi_{b_{0,j}}^{\uparrow,k} +s^k (p^{\uparrow,k}_{b_{0,j}} - p^{\uparrow,k}_{j(b)}), \\
& \lambda_{b_{0,j}}^{k+1} =  \lambda_{b_{0,j}}^{k} +s^k \tilde{h}_{b_0} (g_i^{\text{p},k}, f_{l}^{\text{p},k}, p_{b_{0,j}}^{\uparrow,k}, p_{b_{0,j}}^{\downarrow,k}), \\
& s^{k+1}= \alpha^k s^k  \times \nonumber  \\ &
\times \frac{||\tilde{H} (g_i^{\text{p},k}, f_{l}^{\text{p},k}, p_{b_{0,j}}^{\uparrow,k}, p_{b_{0,j}}^{\downarrow,k}, p_{b(j)}^{\uparrow,k}, p_{b(j)}^{\downarrow,k})||_2}{||\tilde{H} (g_i^{\text{p},k+1}, f_{l}^{\text{p},k+1}, p_{b_{0,j}}^{\uparrow,k+1}, p_{b_{0,j}}^{\downarrow,k+1}, p_{b(j)}^{\uparrow,k+1}, p_{b(j)}^{\downarrow,k+1})||_2}, 
 \end{flalign}
 where $\alpha^k$ is a step-sizing parameter
  \begin{flalign}
& \alpha^k = 1-\frac{1}{M k^{1-1/k^r}}, M>1, r>0.
 \end{flalign}
Value $\tilde{h}_{b_0} (g_i^{p,k}, f_{l}^{p,k}, p_{b_{0,j}}^{\uparrow,k}, p_{b_{0,j}}^{\downarrow,k})$ is defined as the level of constraint violation for a feasible solution of (29) defined for each distribution system as: 

\begin{flalign}
& \tilde{h}_{b_0} (g_i^{p,k}, f_{l}^{p,k}, p_{b_{0,j}}^{\uparrow,k}, p_{b_{0,j}}^{\downarrow,k}) = \sum_{i \in \mathcal{I}_{b_0}} g^{p,k}_i   + \sum_{l|r(l)=b_0} f_l^{p,k} \nonumber \\ & \hspace{1.0cm} - \sum_{l|o(l)=b_0} f^{p,k}_l  + p^{\uparrow,k}_{b_{0,j}}- p^{\downarrow,k}_{b_{0,j}} - 
 L^{\text{p}}_{b_0}. \label{relaxation_eq3}
 \end{flalign}
Accordingly, vector $\tilde{H} (g_i^{\text{p},k}, f_{l}^{\text{p},k}, p_{b_{0,j}}^{\uparrow,k}, p_{b_{0,j}}^{\downarrow,k}, p_{b(j)}^{\uparrow,k}, p_{b(j)}^{\downarrow,k}) $ is the surrogate subgradient direction. Each component of this vector represents the constraint violation of (20) and (25)-(26). 


The procedure described in Step 1-4 repeats until the stopping criteria are satisfied such as CPU time, value of the surrogate subgradient norm, or the duality gap, \cite{TD5}.

\section{Case Study}

\subsection{Illustrative Example} \label{sec:study_illustration}

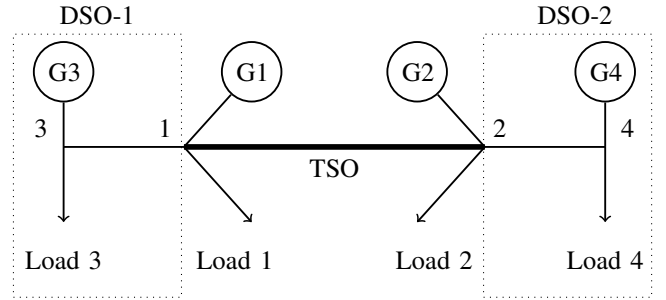
\begin{figure}[!b]
\centering
    \begin{tikzpicture}[auto, node distance=1cm,auto, font=\normalsize]
    \tikzstyle{block} = [circle, draw, fill=white!15, text width=0.4cm, text centered,  minimum height=0.4cm]
    \tikzstyle{block_rect} = [rectangle, draw, dotted, fill=white!10, text width=2.cm, text centered,  minimum height=3.5cm]
    \node[text width=3cm, xshift=-1.5cm] at (-4.25,2.0) {DSO-1};
    \node[text width=3cm, xshift=-0.5cm] at (1.10,2.0) {DSO-2};
    
    \node [block_rect,xshift=-.5cm, yshift=0.0cm)] (dso2) {} ; 
    \node [block_rect,xshift=-6.75cm, yshift=0.0cm)] (dso1) {} ;  
    \node [block, line width=0.25mm, yshift=1.25cm] (gen4) {G4} ;
    \coordinate[below of=gen4] (c4); 
    \coordinate[left of=c4, xshift=-0.6cm] (c2); 

    \node[left of=c4, yshift=.25cm, xshift=-0.4cm] {2}; 
    \node[right of=c4,yshift=.25cm, xshift=-0.7cm] {4};        
    \coordinate[below of=c4] (load4);

    \coordinate[left of=c2, xshift=-3cm] (c1); 
    \coordinate[left of=c1, xshift=-0.6cm] (c3); 
    \node [block, line width=0.25mm, above of = c3] (gen3) {G3} ;
    \coordinate[below of=c3] (load3);  
    \node[right of=c1, yshift=.25cm, xshift=-1.25cm] {1}; 
    \node[left of=c1, yshift=.25cm, xshift=-0.9cm] {3};     
    \node [block, line width=0.25mm, right of = gen3, xshift=1.5cm] (gen1) {G1} ;    
    \node [block, line width=0.25mm, left of = gen4, xshift=-1.5cm] (gen2) {G2} ; 
    \coordinate[below of=gen1, yshift=-1cm] (load1);   
    \coordinate[below of=gen2, yshift=-1cm] (load2);   
	
    \draw [-,thick, line width=0.25mm] (gen1) -- (c1) node [midway] {};
    \draw [-,thick, line width=0.25mm] (gen2) -- (c2) node [midway] {};
    \draw [-,thick, line width=0.25mm] (gen4) -- (c4) node [midway] {};
    \draw [->,thick, line width=0.25mm] (c4) -- (load4) node [near end, below, yshift=-0.5cm] {Load 4};
    \draw [-,thick, line width=0.25mm] (c2) -- (c4) node [near end] {};
    \draw [-,thick, line width=0.75mm] (c2) -- (c1) node [midway] {TSO};
     \draw [-,thick, line width=0.25mm] (c1) -- (c3) node [near end] {};
     \draw [-,thick, line width=0.25mm] (gen3) -- (c3) node [near end] {};  
    \draw [->,thick, line width=0.25mm] (c3) -- (load3) node [near end, below, yshift=-0.5cm] {Load 3};
    \draw [->,thick, line width=0.25mm] (c1) -- (load1) node [near end, below, yshift=-0.5cm] {Load 1};    
    \draw [->,thick, line width=0.25mm] (c2) -- (load2) node [near end, below, yshift=-0.5cm] {Load 2};        
\end{tikzpicture}
\caption{An illustrative example with two distribution systems (DSO-1 and DSO-2) connected to the  transmission system (TSO-1).  }
\label{fig_small_example}
\end{figure}

Fig.~\ref{fig_small_example} describes the illustrative test system. The transmission system includes one transmission line between nodes 1 and 2 with $\overline{F}_{1-2} = 100$ MW. The loads connected directly to the transmission system are $L^{\text{p}}_1=100$ MW and $L^{\text{p}}_2=200$ MW. The operating range of G1 and G2, i.e. the range between their minimum and maximum power outputs, is $\big[ 5, 75 \big]$ MW  and  $\big[ 5, 15 \big]$ MW, respectively,  and their price offers are $C_1^{\text{o}}=\$16$/MW and $C_1^{\text{o}}=\$6$/MW. Each distribution system needs to supply $L^{\text{p}}_3=L^{\text{p}}_4=10$ MW. Generators G3 and G4 have the operating range $\big[ 10, 120 \big]$ MW each with the incremental costs of   $C_3^{\text{o}}=\$6$/MW and $C_4^{\text{o}}=\$4$/MW. For clarity it is assumed that the distribution system has no reactive power loads, as well as power flow and voltage limits.

\begin{figure}[b!]
  \centering
    \includegraphics[width=0.8\linewidth, scale=0.5 ]{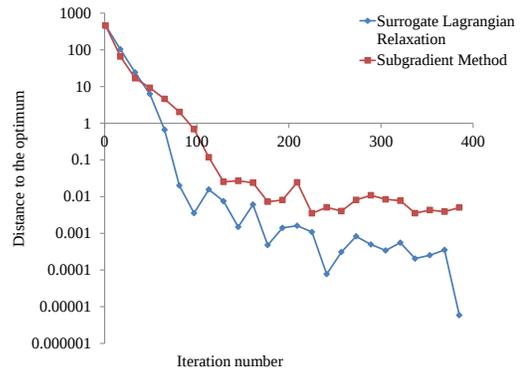}
    \caption{Convergence of the proposed SLR-based approach compared to the convergence of the subgradinet method.}
    \label{fig_convergence_gap}
\end{figure}

\begin{figure}[t!]
  \centering
    \includegraphics[width=0.8\linewidth]{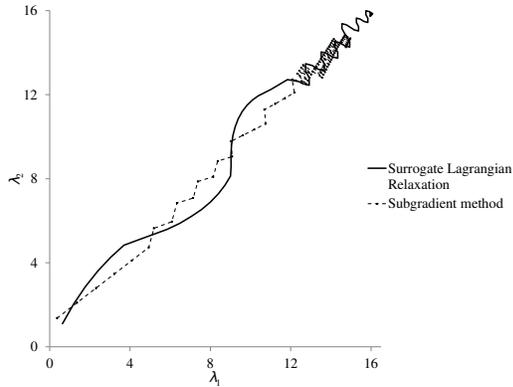}
    \caption{Convergence of the Lagrangian multipliers $\lambda_1$ and $\lambda_2$ (LMPs at node 1 and node 2) to their optimal value of \$16/MW.}
    \vspace{-0.6cm}
    \label{fig_convergence_lamdba}
\end{figure}

The optimal dispatch is $G_1=65$ MW, $G_2=15$ MW, $G_3=120$ MW, and $G_4=120$ MW and the LMPs are $\lambda_1=\lambda_2 = \$16$/MW. Note  G1 is a price-maker as other generators are at their power output limit. The power flow in line between nodes 1 and 2 is 75 MW, and the power flows in distribution lines 1-3 and node 2-4 are 110 MW each. Fig.~\ref{fig_convergence_gap}  compares the convergence of the proposed SLR-based approach observed at each iteration  with the subgradient method, a common algorithmic benchmark. Relative to the benchmark, the proposed approach requires fewer iterations to achieve a higher accuracy of the optimal solution, e.g. after 400 iterations the accuracy gain is roughly 100x. Fig.~\ref{fig_convergence_lamdba} shows how $\lambda_1$ and $\lambda_2$ converge to their optimal values of \$16/MW. As shown in Fig.~\ref{fig_convergence_lamdba}, the SLR reduces zigzaging of Lagrangian multipliers relative to the standard subgradient method, which improves its convergence (Fig.~\ref{fig_convergence_gap}).

\vspace{-0.2cm}
\subsection{IEEE Benchmark} \label{sec:study_ieee}
\vspace{-0.5cm}

\begin{center}
\centering
\vspace{-0.1cm}
\begin{table}[!b]
\centering
\captionsetup{justification=centering, labelsep=period, font=footnotesize, textfont=sc}
\caption{Cost savings and computing times obtained with the proposed TSO-DSO coordination.}
\label{results_ieee_table}
\begin{tabular}{  c | c | c| c  }  
 \hline
 \multirow{ 2}{*}{\# of DSOs} & \multirow{ 2}{*}{\makecell{TSO cost \\  savings, \%}} &\multirow{ 2}{*}{\makecell{DSO* cost \\  savings, \%}} & \multirow{ 2}{*}{CPU time (s)} \\
   & &  &  \\ 
 \hline
 1 & 0.82\% & 0.06\% & 2 \\ 
 2 & 0.82\% & 0.05\%  & 4 \\
 4 & 0.82\% & 0.06\% & 5 \\
 8 & 0.86\% & 0.07\%   & 45 \\
 16 & 1.61\% & 0.19\% & 112 \\
 32 & 3.11\% & 0.18\%  & 234  \\
 64 & 4.29\% & 0.29\%  & 422 \\
 \hline
\end{tabular} \\
* Refers to the total cost of all DSOs cooridnated with the TSO.
\end{table}
\end{center}

This section uses the IEEE 118-bus data \cite{ieee_118} for the transmission system and each distribution system is modeled using the 34-bus IEEE distribution data \cite{ieee_34}. In the following simulations we increase the number of distribution systems connected to the transmission system and compare the results to the case when the TSO and DSO are operated without coordination. When added to the transmission system at a given bus, the distribution system is assumed to fully replace the transmission load at that bus. Each distribution system is assumed to have the same topology and the loads in each distribution system are scaled proportionally to match the total transmission load in the case when the transmission and distribution systems are not coordinated.


Table~\ref{results_ieee_table} summarizes the cost savings obtained with the proposed TSO-DSO coordination, as compared to the case without any coordination, and computing times obtained with the proposed solution technique. As the number of DSOs coordinated with the TSO increases, the  relative TSO and DSO cost savings both increase. However, the  TSO cost savings are roughly one order of magnitude grater than the DSO savings. This observation suggests that the TSO stands to benefit to a larger extent from the proposed coordination and therefore there is a need to design appropriate incentive mechanisms to engage DSOs in the proposed coordination. The computing times also increase with the number of DSOs engaged in the proposed coordination; however, the proposed solution technique is capable of solving all instances within a reasonable amount of time.

\vspace{-0.15cm}
\section{Conclusion \& Future Work}
\vspace{-0.15cm}
This paper presents an model to coordinate transmission and distribution system, while considering binary UC decisions. We solve the proposed model using the Surrogate Lagrangian Relaxation. Our case study demonstrates that both the transmission and distribution systems benefit from the proposed coordination. We also show that the proposed SLR solution technique outperforms existing methods. 

The proposed model points to multiple directions for further investigation.  First, it is important to extend the proposed model to a multi-period framework and include relevant inter-temporal constraints. Extending the model to multiple time periods will also require accounting for demand- and supply-side uncertainty in both the transmission and distribution system. It will also be important to refine the accuracy of AC and DC power flow models used in this work and avoid making restrictive assumptions on the system topology (meshed or radial). Finally, the proposed model and solution technique can be extended to a decentralized decision-making framework to respect  privacy concerns of the DSO and TSO operators.

\end{document}